\documentclass[conference]{IEEEtran}
\usepackage{}
\usepackage{tipa}
\usepackage{bbm}
\usepackage{mathrsfs}
\usepackage{amsfonts}
\usepackage{amssymb,amsmath}
\usepackage{indentfirst}
\usepackage{algorithmic}
\usepackage{algorithm}
\usepackage{enumerate}
\usepackage{stfloats}
\usepackage{amsthm}
\usepackage{cite}
\usepackage{tablefootnote}
\IEEEoverridecommandlockouts
\usepackage{graphicx}
\usepackage{epstopdf}
\usepackage{caption}
\usepackage{float}
\usepackage{wrapfig}
\usepackage{cite,url,subfigure,epsfig,graphicx}
\usepackage{times,verbatim,amsfonts,amsmath,color}

\begin{document}
\title{
Matching Theory for Future Wireless Networks: Fundamentals and Applications
}

\author{\authorblockN{Yunan Gu$^\textbf{1}$, Walid Saad$^\textbf{2}$, Mehdi Bennis$^\textbf{3}$, Merouane Debbah$^\textbf{4}$, and Zhu Han$^\textbf{1}$} \authorblockA{\small
$^\textbf{1}$ Department of Electrical and Computer Engineering, University of Houston, Houston, TX, USA, emails: \url{{ygu6,zhan2}@uh.edu}\\
$^\textbf{2}$ Wireless@VT, Bradley Department of Electrical and Computer Engineering, Virginia Tech, USA, email: \url{walids@vt.edu}\\
$^\textbf{3}$ CWC - Centre for Wireless Communications, University of Oulu, Finland, email: \url{bennis@ee.oulu.fi}\\
$^\textbf{4}$ Mathematical and Algorithmic Sciences Lab, Huawei France R\&D, Paris, France, email: \url{merouane.debbah@huawei.com}\vspace{-1.05cm}
 }%
   \thanks{The work of Yunan Gu and Zhu Han was supported by the U.S. National Science Foundation under Grants CMMI-1434789, ECCS-1405121, CNS-1443917, CNS-1265268, CNS-0953377, and NSFC 61428101. The work of Walid Saad was supported by the U.S. National Science Foundation under Grants CNS-1460316, CNS-1460333, CNS-1443917, ECCS-1405121, CNS-1265268, and  CNS-0953377.}
}
\date{}

\maketitle\thispagestyle{empty}\maketitle\pagestyle{empty}

\begin{abstract}
The emergence of novel wireless networking paradigms such as small cell and cognitive radio networks has forever transformed the way in which wireless systems are operated. In particular, the need for self-organizing solutions to manage the scarce spectral resources has become a prevalent theme in many emerging wireless systems. In this paper, the first comprehensive tutorial on the use of matching theory, a Nobel-prize winning framework, for resource management in wireless networks is developed. To cater for the unique features of emerging wireless networks, a novel, wireless-oriented classification of matching theory is proposed. Then, the key solution concepts and algorithmic implementations of this framework are exposed. Then, the developed concepts are applied in three important wireless networking areas in order to demonstrate the usefulness of this analytical tool. Results show how matching theory can effectively improve the performance of resource allocation in all three applications discussed.
\end{abstract}

%%%%%%%%%%%%%%%%%%%%%%%%%%%%%%%%%%%%%%%%%%%%%%%%%%%%%%%%%%%%%%%
\section{Introduction}\label{sec:intro}
\begin{figure}[!t]
  \begin{center}
    \includegraphics[width=\columnwidth]{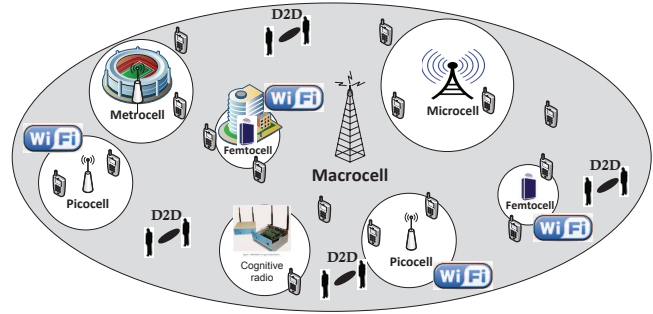}\vspace{-0.2cm}
    \caption{\label{fig:hetnet}  A future wireless network with a mixture of small cells, cognitive radio devices, and heterogeneous spectrum bands. }
   \end{center}\vspace{-1cm}
\end{figure}

%%%%%%%%%%%%%%%%%%%%%%%%%%%%%%%%%%%%%%%%%%%%%%%%%%%%%%%%%%%%%%%
Smartphones, tablets, and other handheld devices are exponentially increasing the traffic load in current wireless networks. To meet this increasing demand, several new paradigms have emerged such as: a) cognitive radio~(CR) networks, in which cognitive devices can adaptively opportunistically access the wireless spectrum thus improving spectral utilization, b) small cell networks that boost wireless capacity and coverage via a viral deployment of low-cost small cell base stations, and c) large-scale device-to-device communications that can occur over both cellular and unlicensed bands. This is gradually leading to a future, multi-tiered heterogeneous wireless architecture, as seen in Fig.~\ref{fig:hetnet}.

Effectively managing resource allocation in such a complex environment warrants a fundamental shift from traditional centralized mechanisms toward self-organizing and self-optimizing approaches. The need for this shift is motivated by practical factors such as the increasing density of wireless networks and the need for communications with low latency.  Even recent emerging centralized paradigms such as cloud-based RAN will still require some form of self-organization due to country-specific backhaul constraints. In consequence, there is a need for self-organizing systems in which small cell base stations and even devices can have some intelligence to rapidly make resource management decisions.

Indeed, there has been a recent surge in literature that proposes new mathematical tools for optimizing resource allocation in many emerging wireless systems. Examples include centralized optimization and game theory. Centralized optimization techniques can provide optimal solutions to resource allocation problems and their algorithmic implementations have matured over the past few years. However, they often require global network information and centralized control thus yielding significant overhead and complexity. This complexity can rapidly increase when dealing with combinatorial, integer programming problems such as channel allocation and user association. Moreover, centralized optimization may not be able to properly handle the challenges of dense and heterogeneous wireless environments such as in Fig. \ref{fig:hetnet}.

The aforementioned limitations of optimization have led to an interesting body of literature that deals with the use of noncooperative game theory for wireless resource allocation~\cite{NSF1}. Despite their potential, such approaches present some shortcomings. First, classical game-theoretic algorithms such as best response will require some form of knowledge on other players' actions, thus limiting their distributed implementation. Second, most game-theoretic solutions, such as the Nash equilibrium, investigate one-sided (or unilateral) stability notions in which equilibrium deviations are evaluated unilaterally per player. Such unilateral deviations may not be practical when investigating assignment problems between distinct sets of players. Last, but not least, the tractability of equilibria in game-theoretic methods requires having some structure in the objective functions which for practical wireless metrics may not always be satisfied.

Recently, \emph{matching theory} has emerged as a promising technique for wireless resource allocation which can overcome some limitations of game theory and optimization~\cite{Jorswieck-11,Leshem-12,Leshem-14,walid-13,Omid-14}. Matching theory is a Nobel-prize winning framework that provides mathematically tractable solutions for the combinatorial problem of matching players in two distinct sets~\cite{MA01,Manlove-13,Irving-87}, depending on the individual information and preference of each player. The advantages of matching theory for wireless resource management include: 1) suitable models for characterizing interactions between heterogeneous nodes, each of which has its own type, objective, and information, 2) ability to define general ``preferences'' that can handle heterogeneous and complex considerations related to wireless quality-of-service (QoS), 3) suitable solutions, in terms of stability and optimality, that accurately reflect different system objectives, and 4) efficient algorithmic implementations that are inherently self-organizing and amenable to fast implementation.

However, reaping the benefits of matching theory for wireless networks requires advancing this framework to handle their intrinsic properties such as interference and delay. Despite the surge in research that applies matching theory for wireless, most existing works are restricted to very limited aspects of resource allocation. This is mainly due to the sparsity of tutorials that tackle
matching theory from an engineering perspective. For instance, most references such as \cite{MA01,Manlove-13,Irving-87} focus on matching problems in microeconomics. In addition, although \cite{seen11} provides an interesting introduction to matching theory for engineering, it does not explicitly explore the challenges of future wireless systems.

In this tutorial, we aim to provide a unified treatment of matching theory oriented towards engineering applications, in general, and wireless networking, in particular. The goal is to gather the state-of-the-art contributions that address the major opportunities and challenges in applying matching theory to the understanding of emerging wireless networks, with emphasis on both new analytical techniques and novel application scenarios. Beyond providing a self-contained tutorial on classical matching concepts, we will introduce a new classification that is oriented towards next-generation wireless systems. For each class of matching problems, we provide the basic challenges, solution concepts, and potential applications. Then, we conclude by summarizing the potential of matching theory as a tool for resource management in wireless networks.

%%%%%%%%%%%%%%%%%%%%%%%%%%%%%%%%%%%%%%%%%%%%%%%%%%%%%%%%%%%%%%%
\section{Matching Theory: Fundamentals and Conventional Classification} \label{sec:fund}
%%%%%%%%%%%%%%%%%%%%%%%%%%%%%%%%%%%%%%%%%%%%%%%%%%%%%%%%%%%%%%%

%%%%%%%%%%%%%%%%%%%%%%%%%%%%%%%%%%%%%%%%%%%%%%%%%%%%%%%%%%%%%%%
\subsection{Basic Matching Definitions}\label{sec:basic}
%%%%%%%%%%%%%%%%%%%%%%%%%%%%%%%%%%%%%%%%%%%%%%%%%%%%%%%%%%%%%%%
The basic wireless resource management problem can be posed as a \emph{matching problem} between resources and users. Depending on the scenario, the resources can be of different abstraction levels, representing base stations, time-frequency chunks, power, or others. Users can be devices, stations, or smartphone applications.  Each user and resource has a quota that defines the maximum number of players with which it can be matched. The main goal of matching is to optimally match resources and users, given their individual, often different objectives and learned information. Each user (resource) builds a ranking of the resources (users) using a \emph{preference relation}. The concept of a preference represents the individual view that each resource or user has on the other set, based on local information. In its basic form, a preference can simply be defined in terms of an objective utility function that quantifies the QoS achieved by a certain resource-user matching. However, a preference is more generic than a utility function in that it can incorporate additional qualitative measures extracted from the information available to users and resources.

A matching is essentially an allocation between resources and users. The basic solution concept for a matching problem is the so-called \emph{two-sided stable matching}. A matching is said to be two-sided \emph{stable}, if and only if there is no \emph{blocking pair} (BP). A BP for a stable marriage case is defined as a pair of user and resource $(u,r)$, where $u$ prefers $r$ to its currently matched user $j$, and $r$ prefers $u$ to its currently matched resource $k$. Thus, $u$ will leave $i$ to be matched to $r$ and $r$ would prefer being matched to user $u$ than user $k$. The implication of stability in a wireless network will be further discussed in Section~\ref{sebsec:wir_solution}. This definition of stability can extend to all types of matching problems.

%%%%%%%%%%%%%%%%%%%%%%%%%%%%%%%%%%%%%%%%%%%%%%%%%%%%%%%%%%%%%%%
\subsection{Conventional Classification}\label{subsec:classical}
%%%%%%%%%%%%%%%%%%%%%%%%%%%%%%%%%%%%%%%%%%%%%%%%%%%%%%%%%%%%%%%
The classical classification of matching problems is based on the values of the player quotas as follows:
\begin {itemize}
\item \emph{One-to-one matching:} Each player can be matched to at most one member of the opposite set. The most prominent example is the stable marriage problem in which men and women need to be matched for marriage.
\item \emph{Many-to-one matching:} Here, in one of the sets, at least one player can be matched to multiple players of the opposing set, while in the other set, every player has exactly one match. One example is the college admissions problem in which one student can be matched to one university while a university can recruit multiple students.
\item \emph{Many-to-many matching:} At least one player within each of the two sets could be matched to more than one member in the other set. Many-to-many matching is the most general type of problems and it admits many examples such as creating partnerships in peer-to-peer networks.
\end {itemize}

There exists other classifications for matching problems, such as based on the partitioning of players, and the preference requirement for players. However, such classes can be often derived as special cases of the above matching problems.

%%%%%%%%%%%%%%%%%%%%%%%%%%%%%%%%%%%%%%%%%%%%%%%%%%%%%%%%%%%%%%%
\subsection{Basic Algorithmic Solution: Deferred Acceptance}\label{subsec:DA}
%%%%%%%%%%%%%%%%%%%%%%%%%%%%%%%%%%%%%%%%%%%%%%%%%%%%%%%%%%%%%%%

\begin{figure}[t] \centering
  {\includegraphics[width=3.2 in, height=2.4 in]{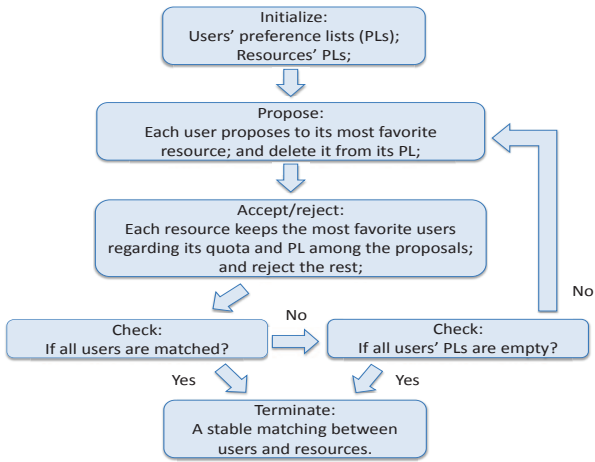}}
 \caption{Deferred acceptance algorithm.}\label{fig:GS}
 \vspace*{-.2in}
\end{figure}

The seminal result in matching theory shows that \emph{at least one stable matching exists} for general preferences in conventional one-to-one and one-to-many games~\cite{Gale-62}. This work also introduced an efficient algorithm, known as the \emph{deferred acceptance~(DA)} algorithm (polynomial time for one-to-one and empirically very fast for one-to-many)  which can find such a matching. DA is an iterative procedure, shown in Fig. \ref{fig:GS}, in which players in one set make proposals to the other set, whose players, in turn, decide to accept or reject these proposals, respecting their quota. Users and resources make their decisions based on their individual preferences (e.g., available information or QoS metric). This process admits many \emph{distributed implementations} which do not require the players to know each other's preferences~\cite{Gale-62}. When the preferences are strict (no indifference), the stable matching is also Pareto optimal for the proposing players~\cite{Gale-62}. Extensions that balance the roles of proposing and receiving players or which handle many-to-many cases have been developed such as in~\cite{seen11} and \cite{walid-14-wnc}.

From an information exchange point of view, even though DA requires players to submit proposals to one another, it does not require a centralized controller. During the information exchange (proposals), each player is required to only collect information on the players they are interested in from the opposite set to perform a ranking according to its preferences. The players need not observe the actions or preferences of other players.
After building preference lists, the players take actions based on the local information they collected without requiring any synchronization in time. The convergence of DA to a stable matching is guaranteed irrespective of the order of play and without any synchronization. Therefore, a DA-based approach can be implemented in a distributed manner without requiring a central information collection center. For such distributed implementations, the results on the polynomial time convergence of one-to-one matching would still hold as corroborated by some recent studies~\cite{Leshem-12,Bayat-12-icc}
%%%%%%%%%%%%%%%%%%%%%%%%%%%%%%%%%%%%%%%%%%%%%%%%%%%%%%%%%%%%%%%
\section{Matching in Wireless Networks: Fundamentals}\label{sec:model}
%%%%%%%%%%%%%%%%%%%%%%%%%%%%%%%%%%%%%%%%%%%%%%%%%%%%%%%%%%%%%%%

%%%%%%%%%%%%%%%%%%%%%%%%%%%%%%%%%%%%%%%%%%%%%%%%%%%%%%%%%%%%%%%
\subsection{Wireless-Oriented Classification}\label{wir_class}
%%%%%%%%%%%%%%%%%%%%%%%%%%%%%%%%%%%%%%%%%%%%%%%%%%%%%%%%%%%%%%%

\begin{figure*}[t]
  \begin{center}
    \includegraphics[width=15cm]{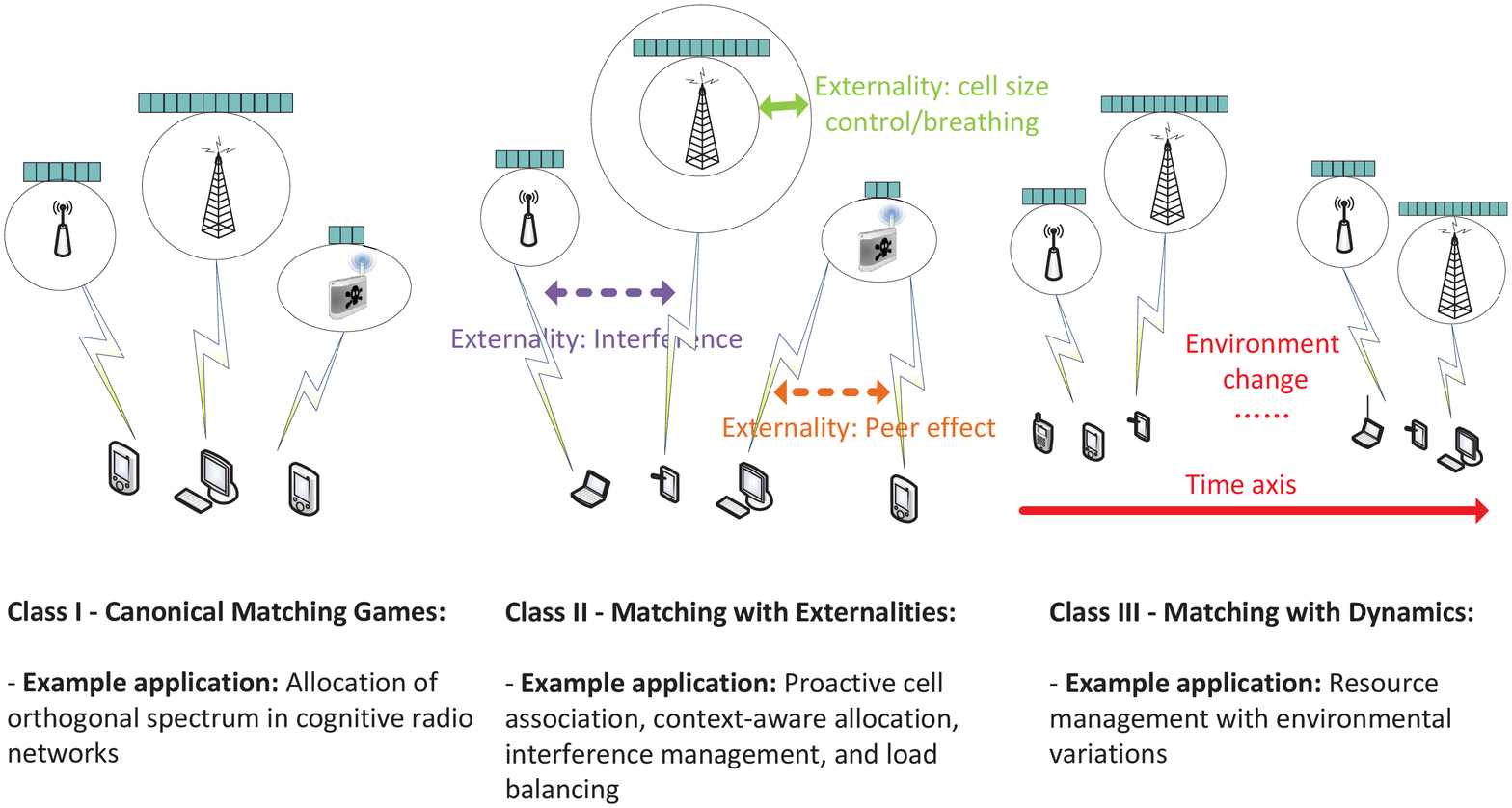}\vspace{-0.3cm}
    \caption{\label{fig:class}  Novel, wireless-oriented classification of matching theory. }
   \end{center}\vspace{-0.7cm}
\end{figure*}

To capture the various wireless resource management features, we condense the rich matching literature into three new, proposed classes of  problems, illustrated in Fig.~\ref{fig:class}, and having the following properties:
\begin{enumerate}
\item  \emph{Class I: Canonical matching:} This constitutes the baseline class in which the preference of any resource (user) depends solely on the information  available at this resource (user) and on the users (resources) to which it is seeking to match. This is useful to study resource management within a single cell or for allocating orthogonal spectrum resources. This is particularly applicable, for example, to CR networks, in which one must allocate orthogonal, licensed channels to a number of unlicensed users.

  \item \emph{Class II: Matching with externalities:}  This class allows finding desirable matchings when the problem exhibits ``externalities'' which translate into interdependencies between the players' preferences. For example, in a small cell network, whenever a user is associated to a resource, the preference of other users will automatically change, since this allocated resource can create interference at other resources using the same frequency. Thus, the preferences of any player depend not only on the information available at this player, but also on the entire matching of the others. We distinguish between two types of externalities: conventional externalities and peer effects. In the former, the dependence of the preferences is between players matched to different players in the other set, such as in the interference example. In the latter, the preference of a user on a resource will depend on the identity and number of other users that are matched to the same resource. Such peer effects are abundant in a wireless environment due to factors such as delay.

\item \emph{Class III: Matching with dynamics:} The third class, matching with dynamics, is suitable for scenarios in which one must adapt the matching processes to dynamics of the environment such as fast fading, mobility, or time-varying traffic. Here, at each time, the preferences of the players might change and, thus, the time dimension must be accounted for in the matching solution. However, for a given time, the matching problem can be of either class I or class II.
\end{enumerate}
Mathematically, the formulation of problems in all three classes will follow the basics of Section~\ref{sec:fund}. For class I, the preferences of one player set simply depend on the other player set. However, for class II, the preferences will now depend not only on the matched user, but also on the entire matching, due to externalities. For class III, one can introduce a time-dependent state variable in the matching. Subsequently, the preferences will now be time and state dependent, if the problem has both dynamics and externalities. The transition between states depends on the application being studied. For example, if the state represents the activity pattern of a licensed user, the transition would follow a classical Markov model. In contrast, if the state represents a dynamically varying fast fading channel, one can use differential equations to represent the state transition.\vspace{-0.2cm}

%%%%%%%%%%%%%%%%%%%%%%%%%%%%%%%%%%%%%%%%%%%%%%%%%%%%%%%%%%%%%%%
\subsection{Matching Theory in Wireless: Discussions}\label{sebsec:wir_solution}
%%%%%%%%%%%%%%%%%%%%%%%%%%%%%%%%%%%%%%%%%%%%%%%%%%%%%%%%%%%%%%%
In wireless resource management, the matching stability notion discussed in Section~\ref{sec:basic} implies robustness to deviations that can benefit both the resource owners and the users. In fact, an unstable matching can for example lead to undesirable cases in which a base station can swap its least preferred user with another since this swap is beneficial to both the resource and the user. Having such network-wide deviations ultimately leads to an unstable network operation. Remarkably, a recent result~\cite{Jorswieck-11} has shown that classical schemes such as proportional fair often yield unstable matchings, which further motivates the need to analyze and optimize stable matchings for self-organizing wireless systems. This concept is very useful in matching problems and is broadly applicable to all classes~\footnote{This concept can also be connected to other stronger or weaker stability notions (e.g., setwise or Nash stability) which have various interpretations.}.

While the existence of a stable matching is guaranteed for canonical games in the one-to-one and one-to-many cases, such results do not readily map to many-to-many nor to classes II and III. However, although DA and its variants were originally conceived for canonical matching, one can also use them as the nexus of new matching algorithms, tailored to the nature of wireless networks. Such algorithms can be used to establish existence of stable matchings for classes II and II as well as for finding outcomes with desirable efficiency properties.

Here, we note that no general existence result for stable matching with externalities exists. However, to handle externalities, one can utilize an iterative DA process which continuously updates the preferences based on the currently perceived matching. By exploiting the structure of externalities via wireless concepts such as interference graphs (e.g., who interferes with whom), one can analyze the convergence and stability of the resulting matching. Naturally, by building on such methods one can expand the realm of matching theory to handle externalities. Similarly, by integrating notions from stochastic games into matching, practical, dynamic algorithms can be devised, to find matchings that can cope with time-varying changes and are stable over time. The basic idea is to cast the matching problem as a stochastic game and, then, explore the rich literature on dynamic game theory~\cite{NSF1} to solve this problem while ensuring that the solution will converge to a \emph{two-sided stable matching} rather than a classical Nash equilibrium. The solution would now essentially be a dynamic and stochastic version of DA.

Although the above discussed matching solutions could provide stable matchings, they also have some limitations: 1) Similar to game theory, matching problems can admit multiple stable solutions and, thus, the selection of a desirable matching is a key design issue, 2) the optimality of the stable solution may not be always guaranteed, however, one can utilize known techniques, such as pricing or optimized utility designs, to drive the matching solution towards an optimal and stable point, 3) the exchange of proposals during DA requires additional signaling in a wireless network, however, one can exploit some structure of the problem to reduce the number of proposals as done in \cite{Leshem-12}.

%%%%%%%%%%%%%%%%%%%%%%%%%%%%%%%%%%%%%%%%%%%%%%%%%%%%%%%%%%%%%%%%
\section{Matching Theory in Wireless Networks: Applications}\label{sec:app}
%%%%%%%%%%%%%%%%%%%%%%%%%%%%%%%%%%%%%%%%%%%%%%%%%%%%%%%%%%%%%%%%

%%%%%%%%%%%%%%%%%%%%%%%%%%%%%%%%%%%%%%%%%%%%%%%%%%%%%%%%%%%%%%%
\subsection{Cognitive Radio Networks}\label{subsec:CRN}
%%%%%%%%%%%%%%%%%%%%%%%%%%%%%%%%%%%%%%%%%%%%%%%%%%%%%%%%%%%%%%%
Cognitive radio networks present a primary application of matching theory due to: 1) necessity of decentralized operation, 2) need for dynamic spectrum access which requires efficient resource management solutions, and 3) the two-sided nature of CR in which licensed, primary users~(PUs) own the channels that must be accessed by the unlicensed, secondary users (SUs). Indeed, in CR networks, centralized optimization solutions are undesirable since PUs and SUs often belong to different operators and cannot be centrally controlled. On the other hand, in some CR problems, such as PU-SU association, using a noncooperative game can lead to stable solutions in which matching PUs and SUs is done without requiring the two-sided consent of both PUs and SUs.

The suitability of matching theory for CR has been corroborated by a number of recent works such as~\cite{Leshem-12} and \cite{Leshem-14}. In particular, \cite{Leshem-12} presented one of the first works in this area. In this work, a one-to-one matching problem is formulated between a number of SUs and a number of PUs (channels). The channels are assumed to be orthogonal and, thus, the game is a canonical matching game. The preferences of both users and channels are based on the same utility function which primarily captures the rate of transmission. Under this model, it is shown that: a) a unique stable matching exists and b) a modified version of the DA algorithm can be used to find the stable allocation in a time efficient manner. This work was extended in \cite{Leshem-14} to account for energy efficiency.

 \begin{figure}[t]
  \begin{center}
    \includegraphics[width=\columnwidth]{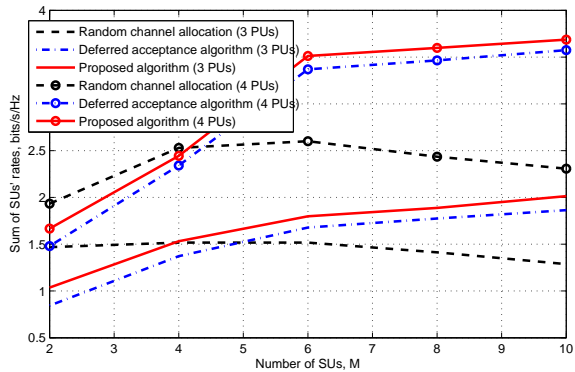}\vspace{-0.2cm}
    \caption{\label{fig:simcr}  Example result showing how matching improves channel allocation in cognitive radio networks. }
   \end{center}\vspace{-1cm}
\end{figure}

Recently, we also studied a one-to-one matching problem between SUs and PUs in which the SUs rank the PUs based on their confidence in sensing the PUs' channels. In particular, using a soft-decision Bayesian framework, we quantified the accuracy of the sensing of each channel and we incorporated this metric in the SUs preferences. Prior to matching, each SU evaluates its appreciation of the PU channel by capturing the effect of confidence in sensing as well as rate. Then, the PUs actively participate in the association process based on two cases: a) when inactive, the PU prefers to grant its band to the  SU with the highest sensing detection and rate to better exploit the channel and b) when active, the PU prefers to protect its band, and, thus, will attempt to limit or deny association. Here, we show that the matching is canonical and we adopt a modified DA algorithm that allows the PUs to handle the aforementioned property. As shown in Fig.~\ref{fig:simcr}, for the studied scenario, the matching-based algorithm yields significant performance gains, in terms of the SUs sum-rate, when compared to classical, random channel allocation schemes. Moreover, the modified DA algorithm also presents sum-rate improvements over classical DA (similar gains can be seen in terms of convergence time).

Clearly, CR networks present an important avenue for matching theory. Many extensions to the existing works can be envisioned, particularly by exploring matching with externalities (under interference constraints) and dynamic matching (given time-varying PU activity).

%%%%%%%%%%%%%%%%%%%%%%%%%%%%%%%%%%%%%%%%%%%%%%%%%%%%%%%%%%%%%%%
\subsection{Heterogeneous Small Cell-based Networks}\label{subsec:hete}
%%%%%%%%%%%%%%%%%%%%%%%%%%%%%%%%%%%%%%%%%%%%%%%%%%%%%%%%%%%%%%%
Heterogeneous small cell-based wireless networks (HetNets) present an important application of matching theory due to their heterogeneity and scale. Also, there has been an increased recent interest in developing context-aware or user-centric HetNets that can exploit new dimensions such as social metrics to improve resource allocation. Such context-awareness further motivates the need for distributed solutions that account for the individual context available at each node -- similar to how matching captures individual preferences. Given this striking analogy between matching theory and resource management in HetNets, the proposed classes can be used to address a variety of problems that include: interference management, handover management, caching, and cell association.

Here, matching is preferred over optimization due to: 1) the density and scale of HetNets which motivate self-organizing solutions, 2) need to account for the context present at each SBS and device instead of a single, global utility function, and 3) the centralized optimization approach will generally yield a combinatorial problem, particularly, in the presence of heterogeneous context, which limits its applicability here. Moreover, although a noncooperative game is also applicable here, it will have a number of limitations that include the need to observe (at least partially) all players preferences and the fact that the solution concepts would not account for two-sided stability as previously explained.

In \cite{walid-14-infocom}, we studied the problem of cell association in the uplink of a HetNet. The basic model here is an uplink HetNet model in which each user needs to decide on which small cell base station (SBS) to be assigned to. The problem is formulated as a one-to-many matching model in which a user can be associated with only one small cell base station (SCBS) and an SCBS can admit a certain quota of users.  The users' preferences over SBSs capture the bit error rate and delay tradeoff that they can achieve while the SBSs' preferences favor load balancing by pushing users to the smaller cells, without jeopardizing QoS. Such a load balancing is essentially a form of cell biasing in which an SBS would offload some users from the macrocell and service them directly. Here, we also consider the delay at each SBS due to the increasing load and the limited capacity of the backhaul that connects the SBS' to the core network. Therefore,  although orthogonal spectrum is considered, due to the delay, the matching problem is shown to have \emph{peer effects} and, thus, it belongs to class II, matching with externalities.

\begin{figure}[t]
  \begin{center}
    \includegraphics[width=\columnwidth]{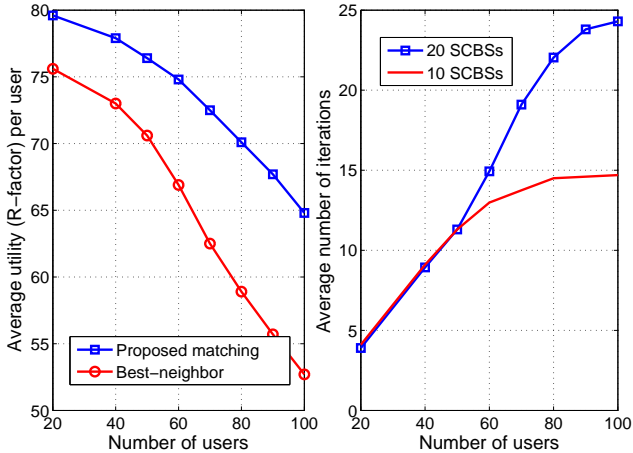}\vspace{-0.2cm}
    \caption{\label{fig:match}  Example result showing how matching theory can be used to improve uplink cell association in HetNets. }
   \end{center}\vspace{-1cm}
\end{figure}

Due to peer effects, applying DA and its variants will not yield a stable matching. Instead, we developed a new
algorithm that starts with a distributed DA-based process using initial preferences based on the worst-case delay. Then, as the nodes measure externalities, they modify their preferences and change their choices by transferring to other SBSs. Then, we show that, due to the presence of transfers in the model, the delay sensitive users will tradeoff two-sided stability for a weaker stability which achieves a better quality-of-service. Fig.~\ref{fig:match} shows simulation results for 2 macrocells and 10 SBSs. We can see that the matching-based approach outperforms the benchmark best neighbor scheme that is often adopted in classical cellular systems, with up to $23\%$ of improvement in the average user utility. Fig.~\ref{fig:match} also shows a reasonable convergence time that grows slowly with the network size.

One can extend this framework of matching with peer effects or, more generally, externalities to many other areas in HetNets. For example, in \cite{walid-13}, the framework is extended to account for interference and perform downlink cell association for a context-aware network in which preferences capture a palette of information that include application type, hardware size, and physical layer metrics. In addition, as shown in in \cite{Bayat-12-icc}, one can explore canonical matching models to study, not only the association at the radio level, but also at the level of operators.

In a nutshell, matching-theoretic models, in all three classes, can serve as a fundamental analytical tool for future cellular systems. Beyond the examples discussed above, one can envision several new models such as many-to-many matching models for caching, dynamic matching models for handling mobility, and stochastic matching models that smartly combine matching with stochastic geometry.

%%%%%%%%%%%%%%%%%%%%%%%%%%%%%%%%%%%%%%%%%%%%%%%%%%%%%%%%%%%%%%%
\subsection{Device-to-Device Communications}\label{subsec:d2d}
%%%%%%%%%%%%%%%%%%%%%%%%%%%%%%%%%%%%%%%%%%%%%%%%%%%%%%%%%%%%%%%
One promising technology to overcome the ever-increasing wireless capacity crunch is device-to-device (D2D) communications. Using D2D, mobile users can communicate directly over cellular spectrum bands while bypassing the base stations~(BSs). As D2D users may share spectral bands with one another as well as with the cellular network, the introduction of D2D in cellular networks will lead to new challenges in terms of interference management and resource allocation. Thus, it will provide an important application for matching theory.

Optimizing resources for D2D communication using centralized optimization can result in more overhead due to information exchange and centralized computation. In particular, centralized optimization will require not only dynamic information collection at the BS from all possible D2D pairs but will also need to deal with an increased computation complexity at the BS level. Formulating the D2D problem as a noncooperative game will be limited by the fact that it will still rely on individual stability and the need for D2D users to observe the preferences of other players. To counter these limitations, it is of interest to develop matching-theoretic models for D2D communications. To this end, we observe that D2D typically involves two types of users: cellular users (CUs) and D2D users (DUs). In the underlay mode of D2D operation, the CUs are exclusively assigned licensed spectrum chunks from the  BSs, while the DUs must share the spectrum with the CUs.

Here, we formulate a two-sided matching problem between the CUs and DUs. Each CU and DU starts by building its preference list by observing the necessary information, e.g., the channel condition, the transmission power, and specific QoS requirements, on the other type of players. Here, the preferences of CUs over DUs are defined as  monetary payment from the DUs or the incurred interference on CUs. The DUs build their preferences over CUs based on the channel conditions or achievable transmission rate. A CU-DU matching is said to be unacceptable if the system's QoS requirements is violated. Players that are unacceptable are then removed from each others preferences. Then, each player (CU or DU) sorts the acceptable players in the descending order of its preferences. After setting up the preferences, proper matching algorithms must be developed to achieve the required system objectives such as maximizing the throughput. For example, when using the DA algorithm, as explained in Fig. \ref{fig:GS}, DUs will propose to the CUs who, in turn, will accept or reject the received applications. The complexity of this iterative process depends on the total number of the acceptable pairs $m$, which is $\mathcal{O}(m)$ \cite{Manlove-13}.

 \begin{figure}[t]
  \begin{center}
    \includegraphics[width=\columnwidth]{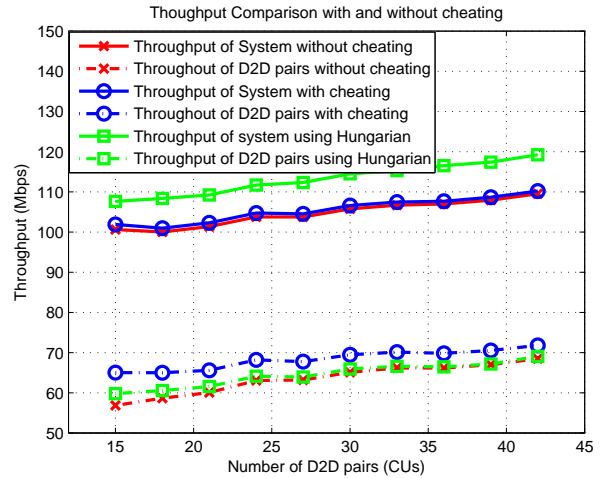}\vspace{-0.2cm}
    \caption{\label{fig:cheat}  Example result showing how cheating further improves both DU and system utilities.}
   \end{center}\vspace{-1cm}
\end{figure}

However, beyond considering a global utility, in some scenarios, one is interested in optimizing the performance of one of the two sets of players, such as the DUs. Based on our work in \cite{Yunan-14}, we proposed the idea of incorporating a form of ``cheating'' in the preferences in an effort to improve the DUs' utilities. Cheating is done by enabling the DUs to smartly change their preferences, so as to reap more performance gains. As shown in our result in Fig.~\ref{fig:cheat}, the use of such cheating strategies can improve the DUs' utility while simultaneously improving the system utility, compared with a DA algorithm. In addition, as done in \cite{Omid-14}, one can extend such D2D models to cases in which D2D communication must explore, beyond physical layer parameters, the social ties of the users. Here, as shown in \cite{Omid-14}, one can cast the problem as a class II problem to capture peer effects which reflect how socially connected are the users who utilize D2D communication with an anchor device (i.e., used for content distribution or caching by the BSs). For such a model, one can enhance the DA algorithm to account for peer effects and show its convergence to a two-sided stable matching.

D2D is undoubtedly an important application area for matching theory with a promising set of future problems.

%%%%%%%%%%%%%%%%%%%%%%%%%%%%%%%%%%%%%%%%%%%%%%%%%%%%%%%%%%%%%%%%
\section{Conclusion}\label{sec:conclusion}
%%%%%%%%%%%%%%%%%%%%%%%%%%%%%%%%%%%%%%%%%%%%%%%%%%%%%%%%%%%%%%%%
In this article, we have provided the first comprehensive tutorial on using matching theory for developing innovative resource management mechanisms in wireless networks. First, we have provided the fundamental concepts of matching theory and discussed a variety of properties that allow the definition of several classes of matching scenarios. Then, we proposed three, new, engineering-oriented classes of matching theory, that can be adopted in wireless networking environments. For each class, we have developed the basic concepts and solutions that can be used to address related problems. We then provided a detailed treatment on how to use such matching-theoretic tools in specific wireless applications. In a nutshell, this paper is expected to provide an accessible and holistic tutorial on the use of new techniques from matching theory for addressing pertinent problems in emerging wireless systems.

\bibliographystyle{IEEEtran}
\bibliography{./tutorial}

\begin{thebibliography}{10}
\providecommand{\url}[1]{#1}
\csname url@samestyle\endcsname
\providecommand{\newblock}{\relax}
\providecommand{\bibinfo}[2]{#2}
\providecommand{\BIBentrySTDinterwordspacing}{\spaceskip=0pt\relax}
\providecommand{\BIBentryALTinterwordstretchfactor}{4}
\providecommand{\BIBentryALTinterwordspacing}{\spaceskip=\fontdimen2\font plus
\BIBentryALTinterwordstretchfactor\fontdimen3\font minus
  \fontdimen4\font\relax}
\providecommand{\BIBforeignlanguage}[2]{{%
\expandafter\ifx\csname l@#1\endcsname\relax
\typeout{** WARNING: IEEEtran.bst: No hyphenation pattern has been}%
\typeout{** loaded for the language `#1'. Using the pattern for}%
\typeout{** the default language instead.}%
\else
\language=\csname l@#1\endcsname
\fi
#2}}
\providecommand{\BIBdecl}{\relax}
\BIBdecl

\bibitem{NSF1}
Z.~Han, D.~Niyato, {W. Saad}, T.~Ba\c{s}ar, and A.~Hj{\o}rungnes, \emph{Game
  Theory in Wireless and Communication Networks: Theory, Models and
  Applications}.\hskip 1em plus 0.5em minus 0.4em\relax Cambridge University
  Press, UK, Oct. 2011.

\bibitem{Jorswieck-11}
E.~A. Jorswieck, ``Stable matchings for resource allocation in wireless
  networks,'' in \emph{Proc. of the 17th International Conference on Digital
  Signal Processing (DSP)}, Corfu, Greece, Jul. 2011.

\bibitem{Leshem-12}
A.~Leshem, E.~Zehavi, and Y.~Yaffe, ``Multichannel opportunistic carrier
  sensing for stable channel access control in cognitive radio systems,''
  \emph{IEEE Journal on Selected Areas in Communications}, vol.~30, no.~1, pp.
  82--95, Jan. 2012.

\bibitem{Leshem-14}
O.~Naparstek, A.~Leshem, and E.~A. Jorswieck, ``Distributed medium access
  control for energy efficient transmission in cognitive radios,'' \emph{arXiv
  preprint arXiv:1401.1671}, 2014.

\bibitem{walid-13}
F.~Pantisano, M.~Bennis, W.~Saad, S.~Valentin, and M.~Debbah, ``Matching with
  externalities for context-aware user-cell association in small cell
  networks,'' Atlanta, GA, Dec. 2013.

\bibitem{Omid-14}
O.~Semiari, W.~Saad, S.~Valentin, and M.~Bennis, ``On self-organizing resource
  allocation for social context-aware small cell networks,'' in \emph{Proc.\
  1st KuVS Workshop on Anticipatory Networks}, Stuttgart, Germany, Sept. 2014.

\bibitem{MA01}
A.~Roth and M.~A.~O. Sotomayor, \emph{Two-Sided Matching: A Study in
  Game-Theoretic Modeling and Analysis}.\hskip 1em plus 0.5em minus 0.4em\relax
  Cambridge University Press, Mar 1992.

\bibitem{Manlove-13}
D.~F. Manlove, \emph{Algorithmics of Matching Under Preferences}.\hskip 1em
  plus 0.5em minus 0.4em\relax World Scientific, 2013.

\bibitem{Irving-87}
R.~W. Irving, P.~Leather, and D.~Gusfield, ``An efficient algorithm for the
  "optimal" stable marriage,'' \emph{Journal of the ACM}, vol.~34, no.~3, pp.
  532--543, Jul. 1987.

\bibitem{seen11}
H.~Xu and B.~Li, ``Seen as stable marriages,'' in \emph{Proc. of the IEEE
  International Conference on Computer Communications (INFOCOM)}, Shanghai,
  China, Mar. 2011.

\bibitem{Gale-62}
D.~Gale and L.~S. Shapley, ``College admissions and the stability of
  marriage,'' \emph{American Mathematical Monthly}, vol.~69, no.~1, pp. 9--15,
  Jan. 1962.

\bibitem{walid-14-wnc}
K.~Hamidouche, W.~Saad, and M.~Debbah, ``Many-to-many matching games for
  proactive social-caching in wireless small cell networks,'' in \emph{Proc.
  12th International Symposium on Modeling and Optimization in Mobile, Ad Hoc,
  and Wireless Networks (WiOpt)}, Hammamet, Tunisia, May 2014.

\bibitem{Bayat-12-icc}
S.~Bayat, R.~H.~Y. Louie, Z.~Han, Y.~Li, and B.~Vucetic, ``Multiple operator
  and multiple femtocell networks: Distributed stable matching,'' in \emph{IEEE
  International Conference on Communications (ICC)}, Ottawa, Canada, Jun. 2012.

\bibitem{walid-14-infocom}
W.~Saad, Z.~Han, R.~Zheng, M.~Debbah, and H.~V. Poor, ``A college admissions
  game for uplink user association in wireless small cell networks,'' in
  \emph{Proc. 33rd Annual IEEE International Conference on Computer
  Communications (INFOCOM)}, Toronto, Canada, Apr.-May 2014.

\bibitem{Yunan-14}
Y.~Gu, Y.~Zhang, M.~Pan, and Z.~Han, ``Cheating in matching of device to device
  pairs in cellular networks,'' in \emph{Proc. of the IEEE Global
  Communications Conference (GLOBECOM)}, Austin, Dec. 2014.

\end{thebibliography}

\end{document}